\documentclass{natureprintstyle}
\usepackage{graphicx}
\usepackage{dcolumn}
\usepackage{bm}
\usepackage{amsmath}
\usepackage{amssymb}
\usepackage{float}
\usepackage{lipsum}
\usepackage{color}
\usepackage{times}
\usepackage{textcomp}
\usepackage{latexsym}
\usepackage[hidelinks]{hyperref}
\usepackage{mathtools}
\usepackage{url}
\usepackage{threeparttable}
\usepackage{xcolor}

\usepackage[utf8]{inputenc}

\begin{document}

\title{Associating binary black holes with galactic centres using lensed gravitational waves}

\author{Laura E. Uronen$^{1,2}$, 
Tian Li$^{3}$, 
Qiuhan He$^{2}$,
Justin Janquart$^{4,5}$, 
Thomas Collett$^{3}$,
Léon V. E. Koopmans$^{2}$,
Otto A. Hannuksela$^{1}$
} 

\maketitle

\begin{affiliations}
\item 
Department of Physics, The Chinese University of
Hong Kong, Shatin, NT, Hong Kong

\item
Kapteyn Astronomical Institute, University of
Groningen, P.O Box 800, 9700 AV Groningen, The
Netherlands

\item
Institute of Cosmology and Gravitation, University of
Portsmouth, Burnaby Rd, Portsmouth, PO1 3FX, UK

\item  
Centre for Cosmology, Particle Physics and Phenomenology - CP3, Université Catholique de Louvain, Louvain-La-Neuve, B-1348, Belgium

\item 
Royal Observatory of Belgium, Avenue Circulaire, 3, 1180 Uccle, Belgium

\end{affiliations}

\smallskip

\maketitle

\begin{abstract}
Binary black holes (BBHs) have been theorised to form through various different channels, but current LIGO-Virgo-KAGRA (LVK)\cite{Aasi2015AdvancedLIGO,Capote2025AdvancedRun,Soni2025LIGORun,Buikema2020SensitivityRun,Tse2019Quantum-EnhancedAstronomy,Martynov2016SensitivityAstronomy,Abbott2016GW150914:Discoveries,Harry2010AdvancedDetectors,Acernese2014AdvancedDetector,Acernese2019IncreasingLight,Acernese2023VirgoRun,Akutsu2021OverviewHistory,Aso2013InterferometerDetector} sky locations make the direct study of formation channels a challenge. 
Association of strongly lensed BBHs to their hosts has been shown to be possible\cite{Hannuksela2020LocalizingLensing} with current detector networks at upgraded sensitivity and survey telescopes like \textit{Euclid}\cite{Wempe2021AObservations}, with localisations that can reach sub-galactic scales in the source plane\cite{Hannuksela2020LocalizingLensing,Wempe2021AObservations,Uronen2024FindingMulti-messenger}. 
With \textit{Hubble Space Telescope}-like high-resolution imaging, we report that strong-lensing localisation of BBHs allows us to confidently associate the BBH with specific structures within their host galaxies, specifically, galactic centres (GCs).
We demonstrate that a BBH localised to the GC of the galaxy can reject light-tracing BBH populations ($p < 5\%$) in a single observation, and in most cases demonstrate high confidence rejection ($p < 1\%$), providing a conservative basis to exclude light-tracing spatial distributions associated with field-like or non-central formation channels. 
Such association therefore directly supports formation channels confined to GCs, such as active galactic nuclei and nuclear stellar clusters.
\end{abstract}

Gravitational-wave (GW) detections have kept increasing in number over observing runs O1, O2, O3, and now O4 from only around a few events in observing run O1 to several hundreds of events in observing run O4\cite{Abbott2016ObservationMerger,Abbott2019GWTC-1:Runs,Abbott2021GWTC-2:Run,Abbott2023GWTC-3:Run,Abac_2025,TheLIGOScientificCollaboration2025GWTC-4.0:Run}, and their number is expected to keep increasing as the sensitivity of the existing detectors improves in observing runs O5 and O6\cite{Gupta2024CharacterizingExplorer}, as more detectors come online, and as next-generation detectors such as the Einstein Telescope and Cosmic Explorer start operating in the 2030s\cite{Gupta2024CharacterizingExplorer,Abac2026TheTelescope}. GWs detected from BBHs whose masses challenge traditional isolated stellar binary evolution theory\cite{Abbott2020GW190521:M_, Abac2025GW231123:M_} have demanded the development of alternative formation theories that could explain how these binaries were formed\cite{Mapelli2021BinaryNOW}: dynamical formation inside dense stellar environments (inside globular clusters or nuclear stellar clusters [NSCs]) or active galactic nuclei (AGN) disks, both of which allow hierarchical formation and are prime candidates to explain these binaries. 
The LVK sky localisations make direct study of environments impossible; for this reason, current efforts to study formation channels typically rely on using simulations to identify signatures left on GW waveforms by their environments\cite{Zwick2025EnvironmentalChannels} or through BBH populations\cite{Sedda2026IsolatedSources}.

\begin{figure}
    \centering
    \includegraphics[width=\linewidth]{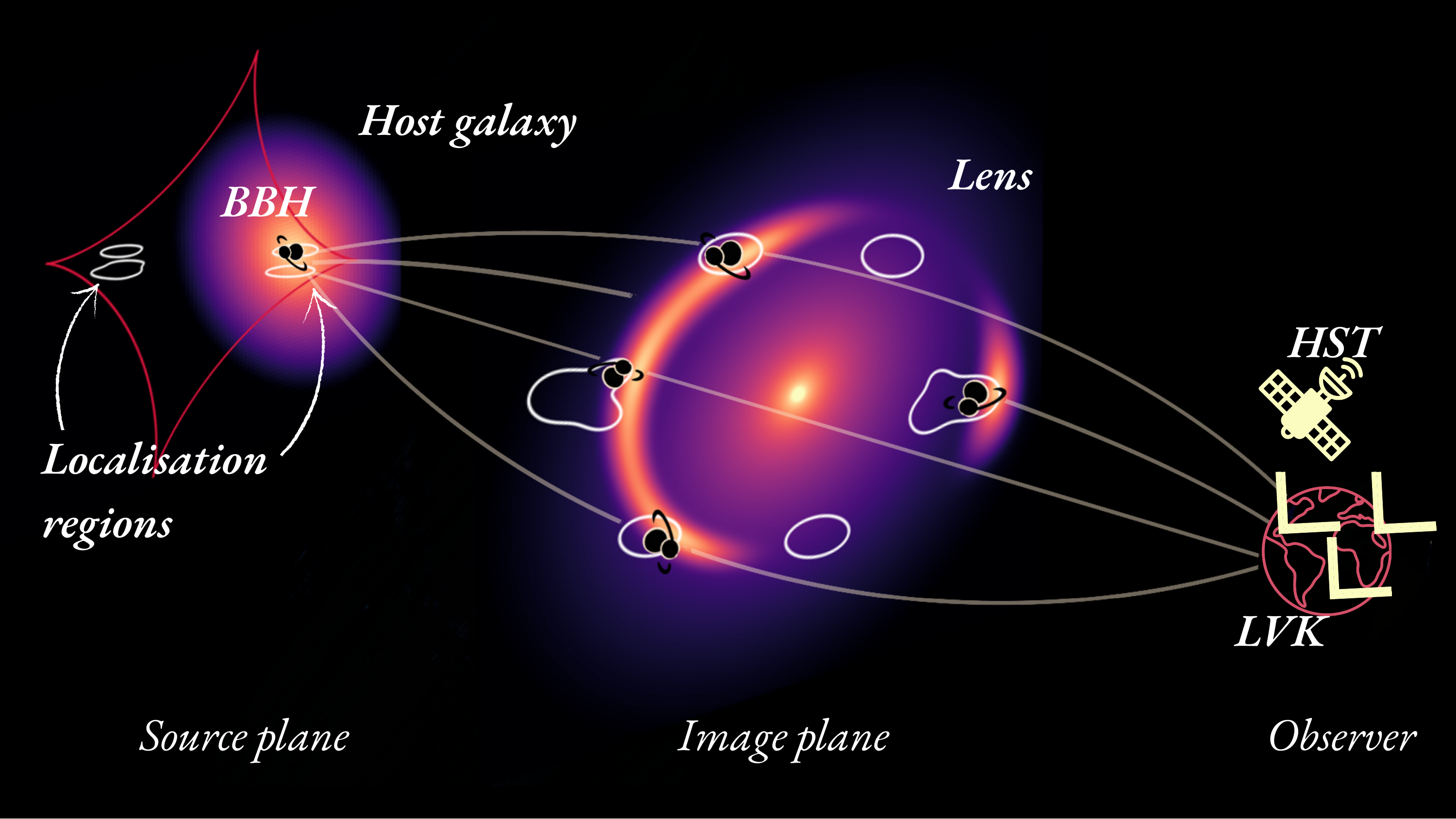}
    \caption{
    Localisation of a BBH originating from the centre of its host galaxy for the large-scale (Einstein radius of 1.8 arcseconds) lens scenario, with the GW localised using a uniform prior. 
    We show the source plane simulation with the BBH at the centre of the host light distribution, and the resulting image plane with lensing. 
    We also display the localisation posteriors, showing the quadrumodal distribution in the source plane resulting from lens symmetries, and the projected image position in the image plane.
    }
    \label{fig:illustration}
\end{figure}

Lensing has also been proposed as a potential solution\cite{Ubach2025Self-lensingDegeneracy,Leong2025ConstrainingWaves}. Several groups have forecasted GW strong lensing detections at a rate of around one in a hundred to a thousand, with the first detections then projected in observing runs O5 and O6\cite{Ng2018PreciseHoles,Li2018GravitationalPerspective,Oguri2018EffectMergers,Xu2022PleasePopulations,Wierda2021BeyondLensing,Phurailatpam2024LerSimulator,Maity2025StrongFuture} and detections becoming common in next-generation detectors\cite{Gupta2024CharacterizingExplorer,Abac2026TheTelescope,Maity2025StrongFuture}, spawning a global hunt for lensing signatures in the LVK data\cite{Hannuksela2019SEARCHEVENTS,McIsaac2020SearchRuns,Dai2020SearchO2,Abbott2021SearchRun,Abbott2024SearchNetwork,Janquart2023Follow-upSearches,TheLIGOScientificCollaboration2025GWTC-4.0:Signatures,Chakraborty2026}, with the current expectation of a few hundred events still being consistent with no detections\cite{Maity2025StrongFuture,Barsode2026}.
A fraction of these events can be localised down to their host galaxies using \textit{Euclid}, \textit{Roman Space Telescope (RST),} \textit{Chinese Space Station Telescope (CSST)} or SKA, and wide-field follow-up supplemented with high-resolution follow-up with instruments such as \textit{HST}, \textit{JWST}, ALMA, \textit{CSST}, or ELT\cite{Hannuksela2020LocalizingLensing,Wempe2021AObservations,Uronen2024FindingMulti-messenger,Shan2025AnWaves}, requiring only the continued development of the GW detector networks and high-resolution electromagnetic follow-up capabilities. 
Direct localisation at sub-arcsecond scales in the source plane using strong lensing\cite{Hannuksela2020LocalizingLensing, Wempe2021AObservations, Uronen2024FindingMulti-messenger, Shan2025AnWaves} leads to the unique possibility of directly associating lensed GWs with structures inside their host galaxies. 
Indeed, population forecasts have shown that formation channels leave distinct imprints on the fraction of lensed events with identifiable lensed hosts and on the stacked spatial distribution of BBHs within those hosts\cite{ChenConstrainingSignals}, though the use of direct spatial localisation within the host galaxy has not yet been studied. 

Within standard population models, BBHs formed in isolation or in globular clusters are expected to preferentially trace stellar populations that are either comparably, or less, centrally concentrated than the overall light distribution of the host galaxy\cite{Lamberts2018PredictingSimulations,Badenes2018StellarView,Belczynski2010OnHoles,Dominik2013DOUBLERATES,Lim2025TheSurveys,Arakelyan2018SpatialGalaxy,Abrusco2025TheCluster} (e.g., older or lower metallicity populations). 
A light-tracing population therefore provides a conservative upper bound on the expected central concentration of BBHs formed through these channels. 
Therefore, if a centrally localised BBH rejects the light-tracing hypothesis $\mathcal{H}_\mathrm{light}$, it also rejects formation channels whose spatial distribution is less skewed towards the GC of the host, and supports through direct empirical localisation formation environments associated with GCs, like NSCs or AGNs. 

We show that with strong lensing localisation, GWs can be directly associated with the host's GC, enabling us to study their origins via direct empirical observation in an unprecedented way. 
Unlike standard approaches that largely rely on population synthesis of BBH formation channels (e.g. N-body or dynamical cluster simulations), our inference relies on observational lens reconstruction and host-galaxy light modelling to reconstruct the source-plane localisation, without assuming a
 specific formation channel prior.
By rejecting light-tracing BBH populations and formation channels associated with these (such as isolated binary formation), the association of the BBH with the GC provides support for central formation environments like AGNs and NSCs. 
We show that for realistic strong-lensing configurations, a BBH localised to the GC can reject a light-tracing population $\mathcal{H}_\mathrm{light}$ with $p < 5\%$ from a single event, and enables empirical discrimination between central and non-central formation channels.

We seek to determine what realistic lensing scenarios can tell us about formation channels. To do this, we simulate three lens configurations consistent with a range of typical, realistic multi-messenger observational scenarios to reflect BBHs with identified EM hosts in the \textit{Euclid} band\cite{Wempe2021AObservations}, using \textsc{lenstronomy}\cite{Birrer2018Lenstronomy:Package}: a small lens ($\theta_E = 0.5"$, $z_l = 1.5, z_s = 2.0$), an average lens ($\theta_E = 1.0", z_l = 0.7, z_s = 1.5$), and a large lens ($\theta_E = 1.8"$, $z_l = 0.5, z_s = 0.8$) (see Table~\ref{tab:results}). All lenses are simulated with an EPL+Shear mass model, standardly used in strong lensing studies. We then simulate our observations with angular resolutions comparable to space-based telescopes like \textit{HST} consistent with high-resolution follow-up of BBH host lenses\cite{Hannuksela2020LocalizingLensing}. To best constrain the GW localisation in the source plane, we simulate only quadruply-imaged lens systems, expected to make up roughly $\sim 30\%$ of aLIGO lensed GW observations\cite{Li2018GravitationalPerspective}.

For the complete multi-messenger (MM) reconstruction of the lens system and the GW localisation to identify the BBH position in the source plane, we first infer the mass model of the lens and brightness profile of the source and lens in the EM band using \textsc{lenstronomy}. We assume in this study the lens and source galaxy that originated the GW have been uniquely identified\cite{Hannuksela2020LocalizingLensing, Wempe2021AObservations}. This identification does depend somewhat on the selected formation channel model, and its correlation with host and localisation properties. However, we expect the general relationship between formation channel, event rates, host galaxy mass, and the probability of a given lens system to host a GW to remain mostly consistent across formation channels, and assume the variation in localisation to be too small to affect the identification probability itself. Future studies will seek to examine the relationship between formation channel models and host identification, and its effect on formation channel association.

We then sample the GW source position in the source plane in conjunction with sampled lens posteriors to reconstruct the GW source position using the time delays and magnifications: 

\begin{align*}
    \log \mathcal{L} &= - \frac{1}{2}\sum^N_{i=1} \left[  
    \left(\frac{\Delta t_{\rm m, i} - \Delta t_{\rm GW, i}}{\sigma_{\Delta t}} \right)^2 + 
    \left( \frac{\mu_{\rm m, i} - \mu_{\rm GW, i}}{\sigma_{\mu}} \right)^2
    \right],
\end{align*}

\noindent where $\Delta t$ are relative time delays, $\mu$ are the relative magnifications, $N$ the number of relative measurements (for 4 GW images, $N=3$), and the subscripts $X_{\rm m}, X_{\rm GW}$ are respectively the values from the sampled model and the measured observations (the full method is detailed in the Methods section and Ref.~\cite{Uronen2026JointHoles}). A demonstration of this localisation is shown in Fig.~\ref{fig:illustration}.

We aim to study whether we can distinguish light-tracing formation channels from central formation channels through either rejection or non-rejection of the light-tracing background $\mathcal{H}_\mathrm{light}$. While this background itself is defined identically across our scenarios, we must also define a suitable prior on that BBH localisation. To do this we use two localisation priors: 

\begin{itemize}
    \item A uniform prior to sample the absolute position of the BBH in the source plane. This provides the most agnostic possible prior on the localisation, and obtains all possible BBH localisation source positions; 
    \item A Gaussian prior to sample the offset of the BBH from the GC at each lens reconstruction posterior. This prior model accounts for two effects: the light-tracing background is also most concentrated at the centre of the galaxy, and thus localisation should account for this by reflecting the central concentration of the light profile probability distribution of the host galaxy, and the expected co-movement of the BBH with the GC position at each lens posterior sample as the GC position is not fixed. 
\end{itemize}

We formally assume that the BBH is, on some level, associated with the host galaxy identified, and fix the redshift to that measured for the host galaxy, rather than assuming the BBH redshift to be plausibly anywhere in the line-of-sight.

We then define a ranking statistic to assess the relationship between the BBH and GC. A natural ranking statistic should therefore relate the position of the BBH localisation to the GC of the host profile, and therefore, we construct a ranking statistic based on the posterior of the relative distance between BBH and GC. We first measure the projected relative separation between the two as
\begin{align}
    r_\mathrm{rel} = \sqrt{(x_\mathrm{GW}-x_\mathrm{GC})^2 + (y_\mathrm{GW}-y_\mathrm{GC})^2 }\,,
\end{align}
where $(x_\mathrm{GW}, y_\mathrm{GW})$ denote the absolute position of the GW inferred through joint EM and GW lens reconstruction and localisation, and $(x_\mathrm{GC}, y_\mathrm{GC})$ are the GC position from the reconstructed source model. For the offset model with a Gaussian prior, we re-cast the sampling parameters as the offset of the BBH from the GC, such that
\begin{align}
    r_\mathrm{rel} = \sqrt{r_\mathrm{x,GW}^2 + r_\mathrm{y,GW}^2 }\,,
\end{align}
where $(r_\mathrm{x,GW}, r_\mathrm{y,GW})$ denote the new GW localisation offset posterior. 

Having computed $r_\mathrm{rel}$ from the posterior of the GW, we define the ranking statistic $r_\mathrm{10}$ we use to assess the proximity of the BBH to the GC.

Strong lensing introduces discrete degeneracies in the source plane localisation caused by lens model symmetries. 
As a result, the $r_\mathrm{rel}$ posterior is multimodal to reflect these symmetries rather than distinct physical hypotheses under a given model. 
As only one of these modes corresponds to the true source position, we focus on quantifying the smallest separation between the GW localisation and GC posterior. We define the ranking statistic $r_{10}$ as the tenth percentile of the $r_\mathrm{rel}$ posterior to provide a conservative estimate of the smallest BBH--GC separations consistent with the data, while remaining insensitive to posterior asymmetries and sampling noise. This is the main statistic used in our study.

$r_\mathrm{10}$ can assess the proximity of a BBH to the GC as a measure of its distance from central galactic structures like AGNs and NSCs. 
However, the value of $r_\mathrm{10}$ alone is not enough to quantify the association of the BBH with the GC: BBHs originating from isolated formation are to first approximation expected to correlate with the mass of the host galaxy\cite{Lamberts2018PredictingSimulations,Badenes2018StellarView,Belczynski2010OnHoles,Dominik2013DOUBLERATES}, and BBHs originating from globular clusters will likely have correlations to the mass profile, though with anisotropic distribution\cite{Lim2025TheSurveys,Arakelyan2018SpatialGalaxy,Abrusco2025TheCluster}. 
The chance that a BBH from a non-central formation channel coincidentally lies close to the GC, where the mass distribution peaks, is therefore non-negligible.

We then proceed to the one-tail $p-$value test. Having reconstructed the source galaxy profile for each scenario, we sample a background of $10^4$ BBHs that trace the light distribution as a proxy of the stellar mass distribution of the source galaxy for our hypothesis testing. Applying this framework, we find that in all configurations the central BBH rejects the light-tracing BBH background, and thus rejects the null hypothesis ($p < 5\%$). 
Our results are presented in Table~\ref{tab:results}.

\begin{figure*}
    \centering
    \includegraphics[width=\linewidth]{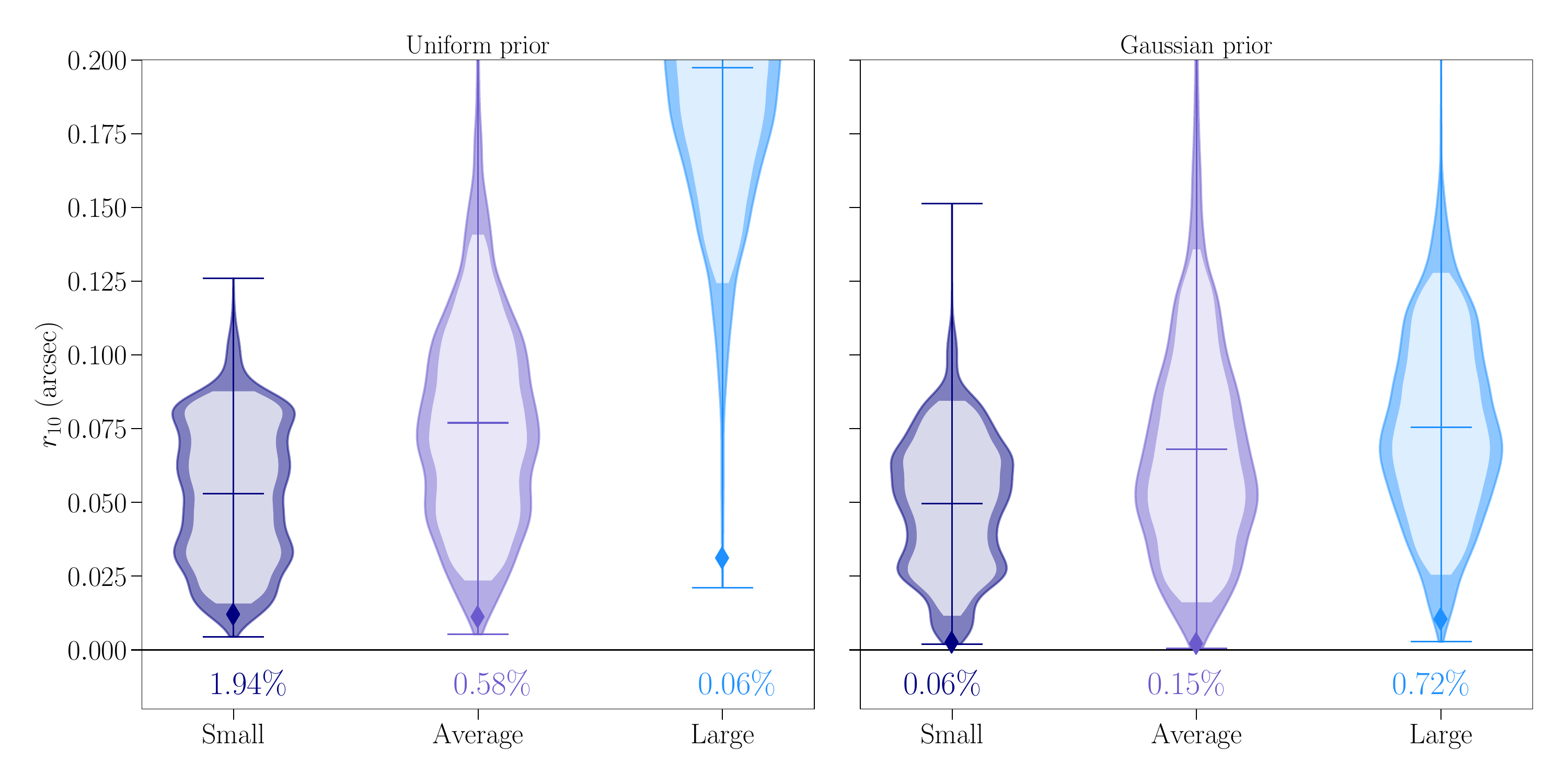}
    \caption{BBH source localisations for the small, average and large lens systems under a uniform prior (left) and Gaussian prior centred on the light distribution (right) (Table~\ref{tab:results}). We localise the BBH then compute the tenth percentile $r_{10}$ of the localisation posterior. The outer violins depict the full background, while the inner regions show the $90\%$ (one-tailed $p<5\%$) confidence interval. The diamond marker shows the central, test BBH, and the false-alarm probability is given beneath each distribution. Every case rejects light-tracing formation channels, thus indicating the BBH originates from the centre of the host galaxy.}
    \label{fig:results}
\end{figure*}

\begin{table}
    \centering
    \begin{tabular}{l  ccc}
         Configuration &  $r_\mathrm{10}$ ($''$) & $r_\mathrm{10}$ (pc) &  $p$ (\%) \\
        \hline
        \hline
        {\it Astrophysical $r_{10}$, $\sim\mathcal{U}$}\\
        \hline
        Small & 0.01208 & 103.64 & 1.94\\
        Average & 0.01125 & 97.68 & 0.58 \\
        Large & 0.03119 & 240.93 & 0.06\\
        \hline
        {\it Astrophysical $r_{10}$, $\sim\mathcal{N}$}\\
        \hline
        Small & 0.00260 & 22.27 & 0.06\\
        Average & 0.00218 & 18.90 & 0.15\\
        Large & 0.01048 & 80.93 & 0.72\\
        \hline
    \end{tabular}
    \caption{Our case studies showing the ranking test statistic $r_{10}$, the representative physical scale of the test statistic, and the $p-$value against the BBH background with the two BBH localisation priors. All cases reject the null hypothesis and support the true BBH originating from the GC.}
    \label{tab:results}
\end{table}

As larger lens systems provide improved EM constraints and fewer degeneracies in the lens reconstruction, improved localisation in the source plane for BBHs allows us to better distinguish BBHs near the GC from the background population. 
Consistent with this, under the uniform prior the smallest lens would not reject the null hypothesis under a more stringent condition (e.g. $p < 1\%$, versus $p = 1.94\%$ obtained), whereas the large scenario unambiguously rejects the background ($p = 0.06\%$).
However, under the Gaussian prior, all three configurations strongly reject the null hypothesis ($p < 1\%$)\footnote{In the Gaussian configuration, while the small and average lens system BBH localisations effectively retrieve a single posterior mode, the large lens still retrieves two degenerate modes, both of which are reasonably within the light profile of the host galaxy. 
Therefore, while the Gaussian prior does reduce incompatible modes, degenerate modes that are plausible are not fully suppressed. 
We further show tests of the effects of this choice of statistic in our Methods discussion.}. 
These results are displayed in Fig.~\ref{fig:results}, displaying the behaviour of the $p-$value for each astrophysical configuration depending on the prior settings used.

Localisation of lensed BBHs can therefore distinguish BBHs consistent with originating in the centre of the host galaxy from a population of light-tracing BBHs. With the light-tracing population acting as a conservative, centrally-skewed background as compared to the expected spatial distribution of BBHs formed in isolation or in globular clusters, rejection of $\mathcal{H}_\mathrm{light}$ rejects non-central formation channels. This presents the first method based on direct spatial localisation of BBHs within their hosts aimed at studying their association with galactic structures and, therefore, their formation channels, without relying on population synthesis simulations. 

Future work should aim to address some of the simplifying assumptions we have made in this work, such as the use of more complex lens and light models, or the effect of incorrect modelling and further high-uncertainty scenarios. Incorporation of information such as the masses and spins from the GWs themselves can further strengthen the statistic, as the probability of formation by a given channel is expected to be strongly correlated with these parameters. This study used only 4-image systems, which are expected to result in the highest probability of correct host identification, and best constrain the GW localisation. Accounting for different numbers of detected lensed GW images as well as the probability of identifying the correct lens will also be critical to study rates and false-alarms of this association. We did not probe the effects of the mass-sheet degeneracy (MSD) on the results. As the mass sheet would affect both the lens reconstruction and GW localisation simultaneously, we do not expect the results regarding the relative localisation of the BBH position to change under the MSD, but future studies should formally establish this.

The study is presented with typical LVK sensitivities and space-based observational resolutions, but with future GW detectors like Cosmic Explorer or the Einstein Telescope, we expect to see hundreds if not thousands of lensed GW detections\cite{Abac2026TheTelescope,Gupta2024CharacterizingExplorer}. A more extensive study based on full galaxy catalogues and a larger variety of lens models would also provide insight into larger, population-scale effects and trends that can result from repeated observations of lensed BBHs associated with GCs, and understand the large-scale effects from future GW detectors.

\section*{Methods}

\paragraph{Simulated observations} Table~\ref{tab:configs} shows the observational settings used, as well as the main settings used for the lensing scenarios (other lens parameters, such as $e_1, e_2$ and shears $\gamma$, are fixed to representative values consistent with typical EPL+Shear systems in the literature, as these parameters have limited impact on source-plane localisation). 

The observational settings used reflect a scenario consistent with \textit{HST-}like resolution, though undersampled, and with a larger pixel scale ($0.067"$) than the \textit{HST} optical WFC or WFC3 pixel scales ($0.05"$ and $0.04"$ respectively). Other \textit{HST} instruments have larger pixel scales ($\sim 0.1"$), and we chose therefore a representative average pixel scale matching the FWHM. Both undersampling and a larger pixel scale lead to a conservative result under non-idealistic observational imaging. Supersampling and smaller pixel scales will improve the imaging, lens reconstruction, and therefore subsequently the GW localisation.

The redshifts for both lens and source are fixed under the assumption that spectroscopic redshift measurements have been obtained for both components. Magnitudes for the source ($m = 25-26$) and lens galaxies ($m = 22$) were chosen to be consistent with distributions in Ref. \cite{Wempe2021AObservations} for joint-EM and GW lens systems to produce observable lens systems. Other values described in Table~\ref{tab:configs} were similarly chosen from distributions in Ref. \cite{Wempe2021AObservations} to reproduce systems matching the intended constraints for each astrophysical scenario. We also present for reference the results of the lens reconstruction for $\theta_E$ and $\gamma$ across systems in Table~\ref{tab:mcmc}. We fix the cosmology to \textsc{Planck18} as implemented in \textsc{astropy}\cite{astropy:2013, astropy:2018, astropy:2022}, and the GW relative magnification uncertainties to $20\%$, consistent with typical GW magnification posterior uncertainties. GWs are generally insensitive to stellar microlensing\cite{cheung2021,yeung2023} and if identified, these can modelled and extracted\cite{zumalacárregui2026effectivedescriptionlensedgravitational,diego2019}---we therefore do not consider additional microlensing effects.

\begin{table}
    \centering
    \begin{tabular}{l c c c c}
        Parameter & Fixed & Small & Average & Large \\
        \hline
        \hline
        {\it Observational settings} & \\
        Pixel scale & 0.067" \\
         FWHM & 0.067"\\
        \hline
        {\it Lens, EPL+Shear} & \\
         Einstein radius, $\theta_E$ & & 0.5" & 1.0" & 1.8"\\
         Lens redshift, $z_l$ & & 1.5 & 0.7 & 0.5\\
        \hline
        {\it Lens light} & \\
        Sérsic radius, $R_s$ & & 0.1"& 0.2"& 0.1" \\
        \hline
        {\it Source galaxy} & \\
        $R_s$ & & 0.02" & 0.05" & 0.02" \\
        $z_s$ & & 2.0 & 1.5 & 0.8\\
        \hline
    \end{tabular}
    \caption{Observation and model settings used for the study in each astrophysical configuration reflecting distributions in Ref.~\cite{Wempe2021AObservations}}
    \label{tab:configs}
\end{table}

\begin{table}
    \centering
    \begin{tabular}{l c c c}
        Parameter & Small & Average & Large \\
        \hline
        \hline
        $\theta_E$ & $0.503\pm0.006$ & $0.998\pm0.007$ & $1.805\pm0.008$ \\
        $\gamma$ & $1.952\pm0.062$ & $1.980\pm0.106$ & $2.000\pm0.109$ \\
    \end{tabular}
    \caption{Lens reconstruction results for each system for the Einstein radius and power-law index, averaged over runs.}
    \label{tab:mcmc}
\end{table}

\paragraph{Time-delay modelling uncertainties}

For a lens system with image position $\Vec\beta$, source position $\Vec\theta$, and lens potential $\psi$, the time delays are given by\cite{wambsgansslensing}

\begin{align*}
    \Vec{\Delta t} = \frac{1 + z_l}{c}\frac{D_lD_s}{D_{ls}}\left( \frac{1}{2} (\Vec\theta - \Vec\beta)^2 - \psi(\Vec\theta)\right),
\end{align*}

\noindent where $D_l, D_s, D_{ls}$ are the angular diameter distances, between lens-observer, source-observer, and lens-source, respectively. Following Ref.~\cite{Uronen2026JointHoles} the full time-delay likelihood can be reduced as follows: 

\begin{align*}
    p(d_{GW} | \Delta t_{EM}) &= \int p(d_{GW}, \Delta t_{GW} | \Delta t_{EM})d\Delta t_{GW}\\
    &=\int p(d_{GW}| \Delta t_{GW} ,\Delta t_{EM})p(\Delta t_{GW} | \Delta t_{EM} )d\Delta t_{GW}\\
    &=\int p(d_{GW}| \Delta t_{GW})p(\Delta t_{GW} | \Delta t_{EM} )d\Delta t_{GW}.\\
\end{align*}

Due to the width of the posterior, the first term can be approximated by a delta function around the median, $\overline{\Delta t}_{GW}$:
\begin{align*}
    p(d_{GW} | \Delta t_{EM}) &=\int \delta(\Delta t_{GW} - \overline{\Delta t}_{GW})p(\Delta t_{GW} | \Delta t_{EM} )d\Delta t_{GW}\\
    &= p(\overline{\Delta t}_{GW} | \Delta t_{EM} )\\
\end{align*}

The final term is the localisation likelihood we use, for which we impose a Gaussian likelihood term\cite{Uronen2026JointHoles}: 
\begin{align*}
    p(\overline{\Delta t}_{GW} | \Delta t_{EM} ) = -0.5\sum_i^N \left[ \frac{\Delta t_{i, EM} - \overline{\Delta t}_{i,GW}}{\sigma_{i}} \right]^2
\end{align*}

\noindent where N is the number of relative time delay measurements, and $\sigma$ the composite uncertainty of measurement and EM modelling uncertainty: $\sigma^2 = \sigma_{GW}^2 + \sigma_{EM}^2$. However, as $\sigma_{EM} \gg \sigma_{GW}$, the resulting uncertainty is $\sigma \sim \sigma_{EM}$. Hence, 
\begin{align*}
    p(\overline{\Delta t}_{GW} | \Delta t_{EM} ) = -0.5\sum_i^N \left[ \frac{\Delta t_{i, EM} - \overline{\Delta t}_{i,GW}}{\sigma_{i,EM}} \right]^2 .
\end{align*}

However, the value of this uncertainty will depend on a wide variety of factors: the lens model chosen, the EM telescope resolution scale, the true underlying mass distribution. Due to GW time-domain precision, GWs are sensitive to mass scales that EM imaging is not; we assume therefore that no EM smooth lens model can reproduce GW time delays, whether due to incorrect modelling or the presence of mass structures the imaging cannot capture. We must therefore impose larger uncertainties than measurement uncertainty alone on the time delays derived from the EM lens. As this study is based on simulations, we know that the lens model used recovers the true underlying mass distribution. However, modelling uncertainty will still appear as a result of the observational effects involved, and attempting to use overly small uncertainties results in sampler non-convergence. In this case, we must therefore choose representative uncertainty values, and study their effect on the results. We elaborate on these more in the following section.

\section*{Robustness tests}

\begin{table}
    \centering
    \begin{tabular}{l  ccc}
         Configuration &  $r_\mathrm{10}$ ($''$) & $r_\mathrm{10}$ (pc) &  $p$ (\%) \\
        \hline
        \hline
        {\it Modelling uncertainties}\\
        \hline
        $1\%$* & 0.01125  & 97.68  & 0.58 \\
        $5\%$ & 0.01209 & 104.96  & 0.55 \\
        $10\%$ &  0.03916 & 339.97 &  7.54 \\
        \hline
        {\it Caustic-uniform background}\\
        \hline
        Uniform prior & 0.01129 & 98.05 &  0.56 \\
        Gaussian prior & 0.00283 & 24.61 & 0.17 \\
        \hline
        {\it Alternative statistic}\\
        \hline
        Weighted mean, $<\overline{r}>$& 0.01670 & 145.02 & 0.36\\
        \hline  
    \end{tabular}
    \caption{Results of additional robustness checks.}
    \label{tab:robustness}
\end{table}

\paragraph{Modelling uncertainties}

For any MM lens reconstruction, the GW time delay reconstruction uncertainty becomes dominated by modelling uncertainties\cite{Uronen2026JointHoles}. We initially assume high confidence in model fidelity to the underlying lens mass distribution to probe the non-modelling dependent performance of this method, and place the time delay modelling uncertainties at $1\%$.

To probe the effect of this choice, we finally choose the average uniform-prior ($\theta_E = 1.0"$) configuration and investigate increased modelling uncertainties at $5\%$ and $10\%$. These reflect similar uncertainty values investigated by Ref.~\cite{Wempe2021AObservations} and a higher end intended to probe an expected failure state of the method with a high uncertainty to demonstrate the effect of low model confidence and poor imaging resolution on the result, and motivate the need for high-resolution follow-up and detailed lens reconstruction.

The $1\%$ and $5\%$ show consistent performance ($p = 0.58\%, 0.55\%$ respectively), indicating that the measurements are limited by the EM lens reconstruction rather than moderate changes in the modelling uncertainty. As expected, the further inflated $10\%$ uncertainty scenario leads to a very clear degradation in the results ($p > 5\%$), such that the null hypothesis is no longer rejected. 

\paragraph{Alternative test statistic}

Depending on the lens configuration, the true location of the BBH may not in truth be in the posterior position closest to the GC. We therefore aim to quantify the effect that this has by re-defining our ranking statistic. Our $r_{10}$ statistic ignores posterior modes far away from the galactic light density, but it also artificially reduces sensitivity to plausible alternative localisations.

We tested the robustness of our test statistic of choice, $r_{10}$, in comparison with alternative definitions of the ranking statistic.

As the BBH localisation is multi-modal, the degeneracy is non-negligible if multiple of the degenerate modes are within the host galaxy's light profile. However, if BBHs are of stellar origin, degenerate modes far from the source light distribution lose importance. We therefore compute a flux-weighted mean as

\begin{align}
    <\overline{r}> = \frac{\sum_{i=1}^{N} r_{rel}^i(x,y)f_i(x,y)}{\sum_{i=1}^{N} f_i},
\end{align}

\noindent where $f_i(x,y)$ is the flux at each posterior sample position and $N$ the number of posterior samples. This provides a complementary measure that accounts for the multimodality of the posterior and the expected correlation of the localisation with the host galaxy profile.

As this statistic accounts for the multimodality of the $r_\mathrm{rel}$ posterior, the value of the test statistic moves away from the GC. However, the $p-$value remains largely unchanged compared to the same scenario with the $r_{10}$ definition, showing that the measurement is robust even to alternative definitions of the ranking statistic. The new statistic also strongly rejects the light-tracing hypothesis as an explanation for the central BBH.

\paragraph{Alternative backgrounds}

We also test an alternative choice of BBH background to examine its effect on our $p-$value. Instead of a background tracing the light profile of the host galaxy, we simply sample the BBHs uniformly around the diamond caustic in the source plane. The difference in background has little effect on the $p-$value: the background is rejected with high confidence in both prior scenarios ($p = 0.56\%$ for uniform prior, $p = 0.17\%$ for the Gaussian prior). As the source galaxy profile occupies a significant fraction of the lens caustic space, the largest variation in background BBH density is observed at the edges of the light distribution of the source galaxy. When sampling from the caustic, we also ensure that all BBHs produce four images, while when sampling from the galactic profile some will produce fewer images and result in poorer localisation. We therefore observe no significant variation in the $p-$value when the majority of the BBH background still lies within the galactic source profile.

\paragraph{Conclusion} 

We show that in all our robustness tests except the explicit failure state with high modelling uncertainties, our experiment result remains largely consistent independent of the choice of test background or test statistic. The biggest observable effect comes from the modelling uncertainties, which can cause the background to no longer be rejected. We therefore demonstrate that this method is mainly dependent on the EM observation and modelling resolution, and it is therefore crucial for lensed BBHs to be followed up with high resolution EM imaging.

\begin{addendum}

\item[Acknowledgements] 
LEU is supported by the Hong Kong PhD Fellowship Scheme (HKPFS) from the Hong Kong Research Grants Council (RGC). LEU and OAH acknowledge support by grants from the Research Grants Council of Hong Kong (Project No. CUHK 14304622, 14307923, and 14307724), the start-up grant from the Chinese University of Hong Kong, and the Direct Grant for Research from the Research Committee of The Chinese University of Hong Kong. The authors are grateful for computational resources provided by the LIGO
Laboratory and supported by the National Science Foundation Grants No. PHY-0757058 and No. PHY-0823459. This material is based upon work supported by NSF’s LIGO Laboratory which is a major facility fully funded by the National Science Foundation.

\item[Author Contributions] 
LEU, TL, TC and OAH conceived the project. All authors contributed to the study design. LEU performed the analyses. All authors contributed to the interpretation of the data. OAH, JJ, TC and LVEK supervised the project and contributed to the conceptual development of the study. LEU wrote the first draft of the manuscript with input from all authors. All authors discussed the results, revised the manuscript and approved the final version.

\item[Correspondence and requests for materials] should be addressed to L.E.U. (email: laura.uronen@gmail.com).

\item[Competing Interests] 
Authors declare no competing interests.

\item[Code availability statement] 
The code and data used are available at \url{https://zenodo.org/doi/10.5281/zenodo.22675771}

\end{addendum}

\bibliographystyle{naturemag}
\bibliography{references}

@article{Wempe2021AObservations,
    author = {Wempe, Ewoud and Koopmans, Léon V E and Wierda, A Renske A C and Hannuksela, Otto A and Van Den Broeck, Chris},
    title = {On the detection and precise localization of merging black holes events through strong gravitational lensing},
    journal = {Monthly Notices of the Royal Astronomical Society},
    volume = {530},
    number = {3},
    pages = {3368-3390},
    year = {2024},
    month = {05},
    issn = {0035-8711},
    doi = {10.1093/mnras/stae1023},
    url = {https://doi.org/10.1093/mnras/stae1023},
}

@article{Aasi2015AdvancedLIGO,
    title = {{Advanced LIGO}},
    year = {2015},
    journal = {Classical and Quantum Gravity},
    author = {Aasi, J. and Abbott, B. P. and Abbott, R. and Abbott, T. and Abernathy, M. R. and Ackley, K. and Adams, C. and Adams, T. and Addesso, P. and Adhikari, R. X. and Adya, V. and Affeldt, C. and Aggarwal, N. and Aguiar, O. D. and Ain, A. and Ajith, P. and Alemic, A. and Allen, B. and Amariutei, D. and Anderson, S. B. and Anderson, W. G. and Arai, K. and Araya, M. C. and Arceneaux, C. and Areeda, J. S. and Ashton, G. and Ast, S. and Aston, S. M. and Aufmuth, P. and Aulbert, C. and Aylott, B. E. and Babak, S. and Baker, P. T. and Ballmer, S. W. and Barayoga, J. C. and Barbet, M. and Barclay, S. and Barish, B. C. and Barker, D. and Barr, B. and Barsotti, L. and Bartlett, J. and Barton, M. A. and Bartos, I. and Bassiri, R. and Batch, J. C. and Baune, C. and Behnke, B. and Bell, A. S. and Bell, C. and Benacquista, M. and Bergman, J. and Bergmann, G. and Berry, C. P.L. and Betzwieser, J. and Bhagwat, S. and Bhandare, R. and Bilenko, I. A. and Billingsley, G. and Birch, J. and Biscans, S. and Biwer, C. and Blackburn, J. K. and Blackburn, L. and Blair, C. D. and Blair, D. and Bock, O. and Bodiya, T. P. and Bojtos, P. and Bond, C. and Bork, R. and Born, M. and Bose, Sukanta and Brady, P. R. and Braginsky, V. B. and Brau, J. E. and Bridges, D. O. and Brinkmann, M. and Brooks, A. F. and Brown, D. A. and Brown, D. D. and Brown, N. M. and Buchman, S. and Buikema, A. and Buonanno, A. and Cadonati, L. and Calder{\'{o}}n Bustillo, J. and Camp, J. B. and Cannon, K. C. and Cao, J. and Capano, C. D. and Caride, S. and Caudill, S. and Cavagli{\`{a}}, M. and Cepeda, C. and Chakraborty, R. and Chalermsongsak, T. and Chamberlin, S. J. and Chao, S. and Charlton, P. and Chen, Y. and Cho, H. S. and Cho, M. and Chow, J. H. and Christensen, N. and Chu, Q. and Chung, S. and Ciani, G. and Clara, F. and Clark, J. A. and Collette, C. and Cominsky, L. and Constancio, M. and Cook, D. and Corbitt, T. R. and Cornish, N. and Corsi, A. and Costa, C. A. and Coughlin, M. W. and Countryman, S. and Couvares, P. and Coward, D. M. and Cowart, M. J. and Coyne, D. C. and Coyne, R. and Craig, K. and Creighton, J. D.E. and Creighton, T. D. and Cripe, J. and Crowder, S. G. and Cumming, A. and Cunningham, L. and Cutler, C. and Dahl, K. and Dal Canton, T. and Damjanic, M. and Danilishin, S. L. and Danzmann, K. and Dartez, L. and Dave, I. and Daveloza, H. and Davies, G. S. and Daw, E. J. and DeBra, D. and Del Pozzo, W. and Denker, T. and Dent, T. and Dergachev, V. and DeRosa, R. T. and DeSalvo, R. and Dhurandhar, S. and Diaz, M. and Di Palma, I. and Dojcinoski, G. and Dominguez, E. and Donovan, F. and Dooley, K. L. and Doravari, S. and Douglas, R. and Downes, T. P. and Driggers, J. C. and Du, Z. and Dwyer, S. and Eberle, T. and Edo, T. and Edwards, M. and Edwards, M. and Effler, A. and Eggenstein, H. B. and Ehrens, P. and Eichholz, J. and Eikenberry, S. S. and Essick, R. and Etzel, T. and Evans, M. and Evans, T. and Factourovich, M. and Fairhurst, S. and Fan, X. and Fang, Q. and Farr, B. and Farr, W. M. and Favata, M. and Fays, M. and Fehrmann, H. and Fejer, M. M. and Feldbaum, D. and Ferreira, E. C. and Fisher, R. P. and Frei, Z. and Freise, A. and Frey, R. and Fricke, T. T. and Fritschel, P. and Frolov, V. V. and Fuentes-Tapia, S. and Fulda, P. and Fyffe, M. and Gair, J. R. and Gaonkar, S. and Gehrels, N. and Gergely, L. and Giaime, J. A. and Giardina, K. D. and Gleason, J. and Goetz, E. and Goetz, R. and Gondan, L. and Gonz{\'{a}}lez, G. and Gordon, N. and Gorodetsky, M. L. and Gossan, S. and Go{\ss}ler, S. and Gr{\"{a}}f, C. and Graff, P. B. and Grant, A. and Gras, S. and Gray, C. and Greenhalgh, R. J.S. and Gretarsson, A. M. and Grote, H. and Grunewald, S. and Guido, C. J. and Guo, X. and Gushwa, K. and Gustafson, E. K. and Gustafson, R. and Hacker, J. and Hall, E. D. and Hammond, G. and Hanke, M. and Hanks, J. and Hanna, C. and Hannam, M. D. and Hanson, J. and Hardwick, T. and Harry, G. M. and Harry, I. W. and Hart, M. and Hartman, M. T. and Haster, C. J. and Haughian, K. and Hee, S. and Heintze, M. and Heinzel, G. and Hendry, M. and Heng, I. S. and Heptonstall, A. W. and Heurs, M. and Hewitson, M. and Hild, S. and Hoak, D. and Hodge, K. A. and Hollitt, S. E. and Holt, K. and Hopkins, P. and Hosken, D. J. and Hough, J. and Houston, E. and Howell, E. J. and Hu, Y. M. and Huerta, E. and Hughey, B. and Husa, S. and Huttner, S. H. and Huynh, M. and Huynh-Dinh, T. and Idrisy, A. and Indik, N. and Ingram, D. R. and Inta, R. and Islas, G. and Isler, J. C. and Isogai, T. and Iyer, B. R. and Izumi, K. and Jacobson, M. and Jang, H. and Jawahar, S. and Ji, Y. and Jim{\'{e}}nez-Forteza, F. and Johnson, W. W. and Jones, D. I. and Jones, R. and Ju, L. and Haris, K. and Kalogera, V. and Kandhasamy, S. and Kang, G. and Kanner, J. B. and Katsavounidis, E. and Katzman, W. and Kaufer, H. and Kaufer, S. and Kaur, T. and Kawabe, K. and Kawazoe, F. and Keiser, G. M. and Keitel, D. and Kelley, D. B. and Kells, W. and Keppel, D. G. and Key, J. S. and Khalaidovski, A. and Khalili, F. Y. and Khazanov, E. A. and Kim, C. and Kim, K. and Kim, N. G. and Kim, N. and Kim, Y. M. and King, E. J. and King, P. J. and Kinzel, D. L. and Kissel, J. S. and Klimenko, S. and Kline, J. and Koehlenbeck, S. and Kokeyama, K. and Kondrashov, V. and Korobko, M. and Korth, W. Z. and Kozak, D. B. and Kringel, V. and Krishnan, B. and Krueger, C. and Kuehn, G. and Kumar, A. and Kumar, P. and Kuo, L. and Landry, M. and Lantz, B. and Larson, S. and Lasky, P. D. and Lazzarini, A. and Lazzaro, C. and Le, J. and Leaci, P. and Leavey, S. and Lebigot, E. O. and Lee, C. H. and Lee, H. K. and Lee, H. M. and Leong, J. R. and Levin, Y. and Levine, B. and Lewis, J. and Li, T. G.F. and Libbrecht, K. and Libson, A. and Lin, A. C. and Littenberg, T. B. and Lockerbie, N. A. and Lockett, V. and Logue, J. and Lombardi, A. L. and Lormand, M. and Lough, J. and Lubinski, M. J. and L{\"{u}}ck, H. and Lundgren, A. P. and Lynch, R. and Ma, Y. and MacArthur, J. and MacDonald, T. and MacHenschalk, B. and MacInnis, M. and MacLeod, D. M. and Maga{\~{n}}a-Sandoval, F. and Magee, R. and Mageswaran, M. and Maglione, C. and Mailand, K. and Mandel, I. and Mandic, V. and Mangano, V. and Mansell, G. L. and M{\'{a}}rka, S. and M{\'{a}}rka, Z. and Markosyan, A. and Maros, E. and Martin, I. W. and Martin, R. M. and Martynov, D. and Marx, J. N. and Mason, K. and Massinger, T. J. and Matichard, F. and Matone, L. and Mavalvala, N. and Mazumder, N. and Mazzolo, G. and McCarthy, R. and McClelland, D. E. and McCormick, S. and McGuire, S. C. and McIntyre, G. and McIver, J. and McLin, K. and McWilliams, S. and Meadors, G. D. and Meinders, M. and Melatos, A. and Mendell, G. and Mercer, R. A. and Meshkov, S. and Messenger, C. and Meyers, P. M. and Miao, H. and Middleton, H. and Mikhailov, E. E. and Miller, A. and Miller, J. and Millhouse, M. and Ming, J. and Mirshekari, S. and Mishra, C. and Mitra, S. and Mitrofanov, V. P. and Mitselmakher, G. and Mittleman, R. and Moe, B. and Mohanty, S. D. and Mohapatra, S. R.P. and Moore, B. and Moraru, D. and Moreno, G. and Morriss, S. R. and Mossavi, K. and Mow-Lowry, C. M. and Mueller, C. L. and Mueller, G. and Mukherjee, S. and Mullavey, A. and Munch, J. and Murphy, D. and Murray, P. G. and Mytidis, A. and Nash, T. and Nayak, R. K. and Necula, V. and Nedkova, K. and Newton, G. and Nguyen, T. and Nielsen, A. B. and Nissanke, S. and Nitz, A. H. and Nolting, D. and Normandin, M. E.N. and Nuttall, L. K. and Ochsner, E. and O'Dell, J. and Oelker, E. and Ogin, G. H. and Oh, J. J. and Oh, S. H. and Ohme, F. and Oppermann, P. and Oram, R. and O'Reilly, B. and Ortega, W. and O'Shaughnessy, R. and Osthelder, C. and Ott, C. D. and Ottaway, D. J. and Ottens, R. S. and Overmier, H. and Owen, B. J. and Padilla, C. and Pai, A. and Pai, S. and Palashov, O. and Pal-Singh, A. and Pan, H. and Pankow, C. and Pannarale, F. and Pant, B. C. and Papa, M. A. and Paris, H. and Patrick, Z. and Pedraza, M. and Pekowsky, L. and Pele, A. and Penn, S. and Perreca, A. and Phelps, M. and Pierro, V. and Pinto, I. M. and Pitkin, M. and Poeld, J. and Post, A. and Poteomkin, A. and Powell, J. and Prasad, J. and Predoi, V. and Premachandra, S. and Prestegard, T. and Price, L. R. and Principe, M. and Privitera, S. and Prix, R. and Prokhorov, L. and Puncken, O. and P{\"{u}}rrer, M. and Qin, J. and Quetschke, V. and Quintero, E. and Quiroga, G. and Quitzow-James, R. and Raab, F. J. and Rabeling, D. S. and Radkins, H. and Raffai, P. and Raja, S. and Rajalakshmi, G. and Rakhmanov, M. and Ramirez, K. and Raymond, V. and Reed, C. M. and Reid, S. and Reitze, D. H. and Reula, O. and Riles, K. and Robertson, N. A. and Robie, R. and Rollins, J. G. and Roma, V. and Romano, J. D. and Romanov, G. and Romie, J. H. and Rowan, S. and R{\"{u}}diger, A. and Ryan, K. and Sachdev, S. and Sadecki, T. and Sadeghian, L. and Saleem, M. and Salemi, F. and Sammut, L. and Sandberg, V. and Sanders, J. R. and Sannibale, V. and Santiago-Prieto, I. and Sathyaprakash, B. S. and Saulson, P. R. and Savage, R. and Sawadsky, A. and Scheuer, J. and Schilling, R. and Schmidt, P. and Schnabel, R. and Schofield, R. M.S. and Schreiber, E. and Schuette, D. and Schutz, B. F. and Scott, J. and Scott, S. M. and Sellers, D. and Sengupta, A. S. and Sergeev, A. and Serna, G. and Sevigny, A. and Shaddock, D. A. and Shahriar, M. S. and Shaltev, M. and Shao, Z. and Shapiro, B. and Shawhan, P. and Shoemaker, D. H. and Sidery, T. L. and Siemens, X. and Sigg, D. and Silva, A. D. and Simakov, D. and Singer, A. and Singer, L. and Singh, R. and Sintes, A. M. and Slagmolen, B. J.J. and Smith, J. R. and Smith, M. R. and Smith, R. J.E. and Smith-Lefebvre, N. D. and Son, E. J. and Sorazu, B. and Souradeep, T. and Staley, A. and Stebbins, J. and Steinke, M. and Steinlechner, J. and Steinlechner, S. and Steinmeyer, D. and Stephens, B. C. and Steplewski, S. and Stevenson, S. and Stone, R. and Strain, K. A. and Strigin, S. and Sturani, R. and Stuver, A. L. and Summerscales, T. Z. and Sutton, P. J. and Szczepanczyk, M. and Szeifert, G. and Talukder, D. and Tanner, D. B. and T{\'{a}}pai, M. and Tarabrin, S. P. and Taracchini, A. and Taylor, R. and Tellez, G. and Theeg, T. and Thirugnanasambandam, M. P. and Thomas, M. and Thomas, P. and Thorne, K. A. and Thorne, K. S. and Thrane, E. and Tiwari, V. and Tomlinson, C. and Torres, C. V. and Torrie, C. I. and Traylor, G. and Tse, M. and Tshilumba, D. and Ugolini, D. and Unnikrishnan, C. S. and Urban, A. L. and Usman, S. A. and Vahlbruch, H. and Vajente, G. and Valdes, G. and Vallisneri, M. and Van Veggel, A. A. and Vass, S. and Vaulin, R. and Vecchio, A. and Veitch, J. and Veitch, P. J. and Venkateswara, K. and Vincent-Finley, R. and Vitale, S. and Vo, T. and Vorvick, C. and Vousden, W. D. and Vyatchanin, S. P. and Wade, A. R. and Wade, L. and Wade, M. and Walker, M. and Wallace, L. and Walsh, S. and Wang, H. and Wang, M. and Wang, X. and Ward, R. L. and Warner, J. and Was, M. and Weaver, B. and Weinert, M. and Weinstein, A. J. and Weiss, R. and Welborn, T. and Wen, L. and Wessels, P. and Westphal, T. and Wette, K. and Whelan, J. T. and Whitcomb, S. E. and White, D. J. and Whiting, B. F. and Wilkinson, C. and Williams, L. and Williams, R. and Williamson, A. R. and Willis, J. L. and Willke, B. and Wimmer, M. and Winkler, W. and Wipf, C. C. and Wittel, H. and Woan, G. and Worden, J. and Xie, S. and Yablon, J. and Yakushin, I. and Yam, W. and Yamamoto, H. and Yancey, C. C. and Yang, Q. and Zanolin, M. and Zhang, Fan and Zhang, L. and Zhang, M. and Zhang, Y. and Zhao, C. and Zhou, M. and Zhu, X. J. and Zucker, M. E. and Zuraw, S. and Zweizig, J.},
    number = {7},
    month = {3},
    pages = {074001},
    volume = {32},
    publisher = {IOP Publishing},
    url = {https://iopscience.iop.org/article/10.1088/0264-9381/32/7/074001},
    doi = {10.1088/0264-9381/32/7/074001},
    issn = {0264-9381},
    arxivId = {1411.4547}
}

@article{Capote2025AdvancedRun,
    title = {{Advanced LIGO detector performance in the fourth observing run}},
    year = {2025},
    journal = {Phys. Rev. D},
    author = {Capote, E. and Jia, W. and Aritomi, N. and Nakano, M. and Xu, V. and Abbott, R. and Abouelfettouh, I. and Adhikari, R. X. and Ananyeva, A. and Appert, S. and Apple, S. K. and Arai, K. and Aston, S. M. and Ball, M. and Ballmer, S. W. and Barker, D. and Barsotti, L. and Berger, B. K. and Betzwieser, J. and Bhattacharjee, D. and Billingsley, G. and Biscans, S. and Blair, C. D. and Bode, N. and Bonilla, E. and Bossilkov, V. and Branch, A. and Brooks, A. F. and Brown, D. D. and Bryant, J. and Cahillane, C. and Cao, H. and Clara, F. and Collins, J. and Compton, C. M. and Cottingham, R. and Coyne, D. C. and Crouch, R. and Csizmazia, J. and Cumming, A. and Dartez, L. P. and Davis, D. and Demos, N. and Dohmen, E. and Driggers, J. C. and Dwyer, S. E. and Effler, A. and Ejlli, A. and Etzel, T. and Evans, M. and Feicht, J. and Frey, R. and Frischhertz, W. and Fritschel, P. and Frolov, V. V. and Fuentes-Garcia, M. and Fulda, P. and Fyffe, M. and Ganapathy, D. and Gateley, B. and Gayer, T. and Giaime, J. A. and Giardina, K. D. and Glanzer, J. and Goetz, E. and Goetz, R. and Goodwin-Jones, A. W. and Gras, S. and Gray, C. and Griffith, D. and Grote, H. and Guidry, T. and Gurs, J. and Hall, E. D. and Hanks, J. and Hanson, J. and Heintze, M. C. and Helmling-Cornell, A. F. and Holland, N. A. and Hoyland, D. and Huang, H. Y. and Inoue, Y. and James, A. L. and Jamies, A. and Jennings, A. and Jones, D. H. and Kabagoz, H. B. and Karat, S. and Karki, S. and Kasprzack, M. and Kawabe, K. and Kijbunchoo, N. and King, P. J. and Kissel, J. S. and Komori, K. and Kontos, A. and Kumar, Rahul and Kuns, K. and Landry, M. and Lantz, B. and Laxen, M. and Lee, K. and Lesovsky, M. and Villarreal, F. Llamas and Lormand, M. and Loughlin, H. A. and Macas, R. and Macinnis, M. and Makarem, C. N. and Mannix, B. and Mansell, G. L. and Martin, R. M. and Mason, K. and Matichard, F. and Mavalvala, N. and Maxwell, N. and McCarrol, G. and McCarthy, R. and McClelland, D. E. and McCormick, S. and McRae, T. and Mera, F. and Merilh, E. L. and Meylahn, F. and Mittleman, R. and Moraru, D. and Moreno, G. and Mullavey, A. and Nelson, T. J.N. and Neunzert, A. and Notte, J. and Oberling, J. and O'Hanlon, T. and Osthelder, C. and Ottaway, D. J. and Overmier, H. and Parker, W. and Patane, O. and Pele, A. and Pham, H. and Pirello, M. and Pullin, J. and Quetschke, V. and Ramirez, K. E. and Ransom, K. and Reyes, J. and Richardson, J. W. and Robinson, M. and Rollins, J. G. and Romel, C. L. and Romie, J. H. and Ross, M. P. and Ryan, K. and Sadecki, T. and Sanchez, A. and Sanchez, E. J. and Sanchez, L. E. and Savage, R. L. and Schaetzl, D. and Schiworski, M. G. and Schnabel, R. and Schofield, R. M.S. and Schwartz, E. and Sellers, D. and Shaffer, T. and Short, R. W. and Sigg, D. and Slagmolen, B. J.J. and Soike, C. and Soni, S. and Srivastava, V. and Sun, L. and Tanner, D. B. and Thomas, M. and Thomas, P. and Thorne, K. A. and Todd, M. R. and Torrie, C. I. and Traylor, G. and Ubhi, A. S. and Vajente, G. and Vanosky, J. and Vecchio, A. and Veitch, P. J. and Vibhute, A. M. and Von Reis, E. R.G. and Warner, J. and Weaver, B. and Weiss, R. and Whittle, C. and Willke, B. and Wipf, C. C. and Wright, J. L. and Yamamoto, H. and Zhang, L. and Zucker, M. E.},
    number = {6},
    month = {3},
    pages = {062002},
    volume = {111},
    publisher = {American Physical Society},
    url = {https://journals.aps.org/prd/abstract/10.1103/PhysRevD.111.062002},
    doi = {10.1103/PhysRevD.111.062002},
    issn = {24700029},
    arxivId = {2411.14607}
}

@article{Harry2010AdvancedDetectors,
       author = {{Harry}, Gregory M. and {LIGO Scientific Collaboration}},
        title = "{Advanced LIGO: the next generation of gravitational wave detectors}",
      journal = {Classical and Quantum Gravity},
         year = 2010,
        month = apr,
       volume = {27},
       number = {8},
          eid = {084006},
        pages = {084006},
          doi = {10.1088/0264-9381/27/8/084006},
       adsurl = {https://ui.adsabs.harvard.edu/abs/2010CQGra..27h4006H}
}

@article{Acernese2014AdvancedDetector,
    title = {{Advanced Virgo: a second-generation interferometric gravitational wave detector}},
    year = {2014},
    journal = {Classical and Quantum Gravity},
    author = {Acernese, F. and Agathos, M. and Agatsuma, K. and Aisa, D. and Allemandou, N. and Allocca, A. and Amarni, J. and Astone, P. and Balestri, G. and Ballardin, G. and Barone, F. and Baronick, J. P. and Barsuglia, M. and Basti, A. and Basti, F. and Bauer, Th S. and Bavigadda, V. and Bejger, M. and Beker, M. G. and Belczynski, C. and Bersanetti, D. and Bertolini, A. and Bitossi, M. and Bizouard, M. A. and Bloemen, S. and Blom, M. and Boer, M. and Bogaert, G. and Bondi, D. and Bondu, F. and Bonelli, L. and Bonnand, R. and Boschi, V. and Bosi, L. and Bouedo, T. and Bradaschia, C. and Branchesi, M. and Briant, T. and Brillet, A. and Brisson, V. and Bulik, T. and Bulten, H. J. and Buskulic, D. and Buy, C. and Cagnoli, G. and Calloni, E. and Campeggi, C. and Canuel, B. and Carbognani, F. and Cavalier, F. and Cavalieri, R. and Cella, G. and Cesarini, E. and Mottin, E. Chassande and Chincarini, A. and Chiummo, A. and Chua, S. and Cleva, F. and Coccia, E. and Cohadon, P. F. and Colla, A. and Colombini, M. and Conte, A. and Coulon, J. P. and Cuoco, E. and Dalmaz, A. and D'Antonio, S. and Dattilo, V. and Davier, M. and Day, R. and Debreczeni, G. and Degallaix, J. and Del{\'{e}}glise, S. and Pozzo, W. Del and Dereli, H. and Rosa, R. De and Fiore, L. Di and Lieto, A. Di and Virgilio, A. Di and Doets, M. and Dolique, V. and Drago, M. and Ducrot, M. and Endrczi, G. and Fafone, V. and Farinon, S. and Ferrante, I. and Ferrini, F. and Fidecaro, F. and Fiori, I. and Flaminio, R. and Fournier, J. D. and Franco, S. and Frasca, S. and Frasconi, F. and Gammaitoni, L. and Garufi, F. and Gaspard, M. and Gatto, A. and Gemme, G. and Gendre, B. and Genin, E. and Gennai, A. and Ghosh, S. and Giacobone, L. and Giazotto, A. and Gouaty, R. and Granata, M. and Greco, G. and Groot, P. and Guidi, G. M. and Harms, J. and Heidmann, A. and Heitmann, H. and Hello, P. and Hemming, G. and Hennes, E. and Hofman, D. and Jaranowski, P. and Jonker, R. J.G. and Kasprzack, M. and K{\'{e}}f{\'{e}}lian, F. and Kowalska, I. and Kraan, M. and Kr{\'{o}}lak, A. and Kutynia, A. and Lazzaro, C. and Leonardi, M. and Leroy, N. and Letendre, N. and Li, T. G.F. and Lieunard, B. and Lorenzini, M. and Loriette, V. and Losurdo, G. and Magazz{\'{u}}, C. and Majorana, E. and Maksimovic, I. and Malvezzi, V. and Man, N. and Mangano, V. and Mantovani, M. and Marchesoni, F. and Marion, F. and Marque, J. and Martelli, F. and Martellini, L. and Masserot, A. and Meacher, D. and Meidam, J. and Mezzani, F. and Michel, C. and Milano, L. and Minenkov, Y. and Moggi, A. and Mohan, M. and Montani, M. and Morgado, N. and Mours, B. and Mul, F. and Nagy, M. F. and Nardecchia, I. and Naticchioni, L. and Nelemans, G. and Neri, I. and Neri, M. and Nocera, F. and Pacaud, E. and Palomba, C. and Paoletti, F. and Paoli, A. and Pasqualetti, A. and Passaquieti, R. and Passuello, D. and Perciballi, M. and Petit, S. and Pichot, M. and Piergiovanni, F. and Pillant, G. and Piluso, A. and Pinard, L. and Poggiani, R. and Prijatelj, M. and Prodi, G. A. and Punturo, M. and Puppo, P. and Rabeling, D. S. and R{\'{a}}cz, I. and Rapagnani, P. and Razzano, M. and Re, V. and Regimbau, T. and Ricci, F. and Robinet, F. and Rocchi, A. and Rolland, L. and Romano, R. and Rosi{\'{n}}ska, D. and Ruggi, P. and Saracco, E. and Sassolas, B. and Schimmel, F. and Sentenac, D. and Sequino, V. and Shah, S. and Siellez, K. and Straniero, N. and Swinkels, B. and Tacca, M. and Tonelli, M. and Travasso, F. and Turconi, M. and Vajente, G. and Van Bakel, N. and Van Beuzekom, M. and Van Den Brand, J. F.J. and Van Den Broeck, C. and Van Der Sluys, M. V. and Van Heijningen, J. and Vas{\'{u}}th, M. and Vedovato, G. and Veitch, J. and Verkindt, D. and Vetrano, F. and Vicer{\'{e}}, A. and Vinet, J. Y. and Visser, G. and Vocca, H. and Ward, R. and Was, M. and Wei, L. W. and Yvert, M. and Zny, A. Zadro and Zendri, J. P.},
    number = {2},
    month = {12},
    pages = {024001},
    volume = {32},
    publisher = {IOP Publishing},
    url = {https://iopscience.iop.org/article/10.1088/0264-9381/32/2/024001},
    doi = {10.1088/0264-9381/32/2/024001},
    issn = {0264-9381},
    arxivId = {1408.3978}
}

@article{Shan2025AnWaves,
    title = {{An interference-based method for the detection of strongly lensed gravitational waves}},
    year = {2025},
    journal = {Nature Astronomy},
    author = {Shan, Xikai and Hu, Bin and Chen, Xuechun and Cai, Rong-Gen},
    month = {6},
    pages = {916--924},
    volume = {9},
    doi = {10.1038/s41550-025-02519-5},
    arxivId = {2301.06117}
}

@ARTICLE{Wierda2021BeyondLensing,
       author = {{Wierda}, A. Renske A.~C. and {Wempe}, Ewoud and {Hannuksela}, Otto A. and {Koopmans}, L{\'e}on V.~E. and {Van Den Broeck}, Chris},
        title = "{Beyond the Detector Horizon: Forecasting Gravitational-Wave Strong Lensing}",
      journal = {ApJ},
         year = 2021,
        month = nov,
       volume = {921},
       number = {2},
          eid = {154},
        pages = {154},
          doi = {10.3847/1538-4357/ac1bb4},
archivePrefix = {arXiv},
       eprint = {2106.06303},
 primaryClass = {astro-ph.HE},
       adsurl = {https://ui.adsabs.harvard.edu/abs/2021ApJ...921..154W}
}

@ARTICLE{Mapelli2021BinaryNOW,
       author = {{Mapelli}, Michela},
        title = "{Binary black hole mergers: formation and populations}",
      journal = {Front. Astron. Space Sci.},
         year = 2020,
        month = jul,
       volume = {7},
          eid = {38},
        pages = {38},
          doi = {10.3389/fspas.2020.00038},
archivePrefix = {arXiv},
       eprint = {2105.12455},
 primaryClass = {astro-ph.HE},
       adsurl = {https://ui.adsabs.harvard.edu/abs/2020FrASS...7...38M}
}

@article{Gupta2024CharacterizingExplorer,
    title = {{Characterizing gravitational wave detector networks: from A{\#} to cosmic explorer}},
    year = {2024},
    journal = {Classical and Quantum Gravity},
    author = {Gupta, Ish and Afle, Chaitanya and Arun, K.~G. and Bandopadhyay, Ananya and Baryakhtar, Masha and Biscoveanu, Sylvia and Borhanian, Ssohrab and Broekgaarden, Floor and Corsi, Alessandra and Dhani, Arnab and Evans, Matthew and Hall, Evan D and Hannuksela, Otto A and Kacanja, Keisi and Kashyap, Rahul and Khadkikar, Sanika and Kuns, Kevin and Li, Tjonnie G.~F. and Miller, Andrew L and Harvey Nitz, Alexander and Owen, Benjamin J and Palomba, Cristiano and Pearce, Anthony and Phurailatpam, Hemantakumar and Rajbhandari, Binod and Read, Jocelyn and Romano, Joseph D and Sathyaprakash, Bangalore S and Shoemaker, David H and Singh, Divya and Vitale, Salvatore and Barsotti, Lisa and Berti, Emanuele and Cahillane, Craig and Chen, Hsin-Yu and Fritschel, Peter and Haster, Carl-Johan and Landry, Philippe and Lovelace, Geoffrey and McClelland, David and J J Slagmolen, Bram and R Smith, Joshua and Soares-Santos, Marcelle and Sun, Ling and Tanner, David and Yamamoto, Hiro and Zucker, Michael},
    number = {24},
    month = {12},
    pages = {245001},
    volume = {41},
    doi = {10.1088/1361-6382/ad7b99},
    arxivId = {2307.10421}
}

@article{Leong2025ConstrainingWaves,
    title = {{Constraining Binary Mergers in Active Galactic Nuclei Disks Using the Nonobservation of Lensed Gravitational Waves}},
    year = {2025},
    journal = {ApJL},
    author = {Leong, Samson H W and Janquart, Justin and Sharma, Aditya Kumar and Martens, Paul and Ajith, Parameswaran and Hannuksela, Otto A},
    number = {2},
    month = {5},
    pages = {L27},
    volume = {979},
    publisher = {IOP Publishing},
    url = {https://iopscience.iop.org/article/10.3847/2041-8213/ad9ead},
    doi = {10.3847/2041-8213/AD9EAD},
    issn = {2041-8205}
}

@ARTICLE{ChenConstrainingSignals,
       author = {{Chen}, Zhiwei and {Lu}, Youjun and {Zhao}, Yuetong},
        title = "{Constraining the Origin of Stellar Binary Black Hole Mergers by Detections of Their Lensed Host Galaxies and Gravitational Wave Signals}",
      journal = {Astrophys. J.},
         year = 2022,
        month = nov,
       volume = {940},
       number = {1},
          eid = {17},
        pages = {17},
          doi = {10.3847/1538-4357/ac98b7},
archivePrefix = {arXiv},
       eprint = {2210.09892},
 primaryClass = {astro-ph.HE},
       adsurl = {https://ui.adsabs.harvard.edu/abs/2022ApJ...940...17C}
}

@article{Dominik2013DOUBLERATES,
    title = {{DOUBLE COMPACT OBJECTS. II. COSMOLOGICAL MERGER RATES}},
    year = {2013},
    journal = {ApJ},
    author = {Dominik, Michal and Belczynski, Krzysztof and Fryer, Christopher and Holz, Daniel E. and Berti, Emanuele and Bulik, Tomasz and Mandel, Ilya and O'Shaughnessy, Richard},
    number = {1},
    month = {11},
    pages = {72},
    volume = {779},
    publisher = {IOP Publishing},
    url = {https://iopscience.iop.org/article/10.1088/0004-637X/779/1/72},
    doi = {10.1088/0004-637X/779/1/72},
    issn = {0004-637X},
    arxivId = {1308.1546}
}

@ARTICLE{Oguri2018EffectMergers,
       author = {{Oguri}, Masamune},
        title = "{Effect of gravitational lensing on the distribution of gravitational waves from distant binary black hole mergers}",
      journal = {MNRAS},
         year = 2018,
        month = nov,
       volume = {480},
       number = {3},
        pages = {3842-3855},
          doi = {10.1093/mnras/sty2145},
archivePrefix = {arXiv},
       eprint = {1807.02584},
 primaryClass = {astro-ph.CO},
       adsurl = {https://ui.adsabs.harvard.edu/abs/2018MNRAS.480.3842O}
}

@article{Zwick2025EnvironmentalChannels,
    title = {{Environmental Effects in Stellar Mass Gravitational-wave Sources. I. Expected Fraction of Signals with Significant Dephasing in the Dynamical and Active Galactic Nucleus Channels}},
    year = {2025},
    journal = {ApJ},
    author = {Zwick, Lorenz and Tak{\'{a}}tsy, János and Saini, Pankaj and Hendriks, Kai and Samsing, Johan and Tiede, Christopher and Rowan, Connar and Trani, Alessandro A.},
    number = {2},
    month = {9},
    pages = {131},
    volume = {991},
    publisher = {IOP Publishing},
    url = {https://iopscience.iop.org/article/10.3847/1538-4357/adf6b8},
    doi = {10.3847/1538-4357/ADF6B8},
    issn = {0004-637X}
}

@ARTICLE{Uronen2024FindingMulti-messenger,
       author = {{Uronen}, Laura Elina and {Li}, Tian and {Janquart}, Justin and {Phurailatpam}, Hemanta and {Poon}, Jason and {Wempe}, Ewoud and {Koopmans}, Leon and {Hannuksela}, Otto},
        title = "{Finding black holes: an unconventional multi-messenger}",
      journal = {Phil. Trans. R. Soc. A},
         year = 2025,
        month = apr,
       volume = {383},
       number = {2294},
          eid = {20240152},
        pages = {20240152},
          doi = {10.1098/rsta.2024.0152},
archivePrefix = {arXiv},
       eprint = {2406.14257},
 primaryClass = {astro-ph.HE},
       adsurl = {https://ui.adsabs.harvard.edu/abs/2025RSPTA.38340152U}
}

@ARTICLE{Janquart2023Follow-upSearches,
       author = {{Janquart}, J. and {Wright}, M. and {Goyal}, S. and {Chan}, J.~C.~L. and {Ganguly}, A. and {Garr{\'o}n}, {\'A}. and {Keitel}, D. and {Li}, A.~K.~Y. and {Liu}, A. and {Lo}, R.~K.~L. and {Mishra}, A. and {More}, A. and {Phurailatpam}, H. and {Prasia}, P. and {Ajith}, P. and {Biscoveanu}, S. and {Cremonese}, P. and {Cudell}, J.~R. and {Ezquiaga}, J.~M. and {Garcia-Bellido}, J. and {Hannuksela}, O.~A. and {Haris}, K. and {Harry}, I. and {Hendry}, M. and {Husa}, S. and {Kapadia}, S. and {Li}, T.~G.~F. and {Maga{\~n}a Hernandez}, I. and {Mukherjee}, S. and {Seo}, E. and {Van Den Broeck}, C. and {Veitch}, J.},
        title = "{Follow-up analyses to the O3 LIGO-Virgo-KAGRA lensing searches}",
      journal = {MNRAS},
         year = 2023,
        month = dec,
       volume = {526},
       number = {3},
        pages = {3832-3860},
          doi = {10.1093/mnras/stad2909},
archivePrefix = {arXiv},
       eprint = {2306.03827},
 primaryClass = {gr-qc},
       adsurl = {https://ui.adsabs.harvard.edu/abs/2023MNRAS.526.3832J}
}

@ARTICLE{Li2018GravitationalPerspective,
       author = {{Li}, Shun-Sheng and {Mao}, Shude and {Zhao}, Yuetong and {Lu}, Youjun},
        title = "{Gravitational lensing of gravitational waves: a statistical perspective}",
      journal = {MNRAS},
         year = 2018,
        month = may,
       volume = {476},
       number = {2},
        pages = {2220-2229},
          doi = {10.1093/mnras/sty411},
archivePrefix = {arXiv},
       eprint = {1802.05089},
 primaryClass = {astro-ph.CO},
       adsurl = {https://ui.adsabs.harvard.edu/abs/2018MNRAS.476.2220L}
}

@article{Abbott2020GW190521:M_,
    title = {{GW190521: A Binary Black Hole Merger with a Total Mass of 150 $M_{\odot}$}},
    year = {2020},
    journal = {Phys. Rev. Lett.},
    author = {Abbott, B P and {others}},
    pages = {101102},
    volume = {125},
    number = {10},
    doi = {10.1103/PhysRevLett.125.101102}
}

@article{Abac2025GW231123:M_,
    title = {{GW231123: A Binary Black Hole Merger with Total Mass between 190 and 265 $M_{\odot}$}},
    year = {2025},
    journal = {ApJL},
    author = {Abac, A G and {others}},
    pages = {L25},
    volume = {993},
    doi = {10.3847/2041-8213/ae0c9c}
}

@article{Abbott2019GWTC-1:Runs,
    title = {{GWTC-1: A Gravitational-Wave Transient Catalog of Compact Binary Mergers Observed by LIGO and Virgo during the First and Second Observing Runs}},
    year = {2019},
    journal = {Phys. Rev. X},
    author = {Abbott, B.~P. and Abbott, R and Abbott, T.~D. and Abraham, S and Acernese, F and Ackley, K and Adams, C and Adhikari, R.~X. and Adya, V.~B. and Affeldt, C and Agathos, M and Agatsuma, K and Aggarwal, N and Aguiar, O.~D. and Aiello, L and Ain, A and Ajith, P and Allen, G and Allocca, A and Aloy, M.~A. and Altin, P.~A. and Amato, A and Ananyeva, A and Anderson, S.~B. and Anderson, W.~G. and Angelova, S.~V. and Antier, S and Appert, S and Arai, K and Araya, M.~C. and Areeda, J.~S. and Ar{\`{e}}ne, M and Arnaud, N and Arun, K.~G. and Ascenzi, S and Ashton, G and Aston, S.~M. and Astone, P and Aubin, F and Aufmuth, P and AultONeal, K and Austin, C and Avendano, V and Avila-Alvarez, A and Babak, S and Bacon, P and Badaracco, F and Bader, M.~K.~M. and Bae, S and Baker, P.~T. and Baldaccini, F and Ballardin, G and Ballmer, S.~W. and Banagiri, S and Barayoga, J.~C. and Barclay, S.~E. and Barish, B.~C. and Barker, D and Barkett, K and Barnum, S and Barone, F and Barr, B and Barsotti, L and Barsuglia, M and Barta, D and Bartlett, J and Bartos, I and Bassiri, R and Basti, A and Bawaj, M and Bayley, J.~C. and Bazzan, M and B{\'{e}}csy, B and Bejger, M and Belahcene, I and Bell, A.~S. and Beniwal, D and Berger, B.~K. and Bergmann, G and Bernuzzi, S and Bero, J.~J. and Berry, C.~P.~L. and Bersanetti, D and Bertolini, A and Betzwieser, J and Bhandare, R and Bidler, J and Bilenko, I.~A. and Bilgili, S.~A. and Billingsley, G and Birch, J and Birney, R and Birnholtz, O and Biscans, S and Biscoveanu, S and Bisht, A and Bitossi, M and Bizouard, M.~A. and Blackburn, J.~K. and Blackman, J and Blair, C.~D. and Blair, D.~G. and Blair, R.~M. and Bloemen, S and Bode, N and Boer, M and Boetzel, Y and Bogaert, G and Bondu, F and Bonilla, E and Bonnand, R and Booker, P and Boom, B.~A. and Booth, C.~D. and Bork, R and Boschi, V and Bose, S and Bossie, K and Bossilkov, V and Bosveld, J and Bouffanais, Y and Bozzi, A and Bradaschia, C and Brady, P.~R. and Bramley, A and Branchesi, M and Brau, J.~E. and Briant, T and Briggs, J.~H. and Brighenti, F and Brillet, A and Brinkmann, M and Brisson, V and Brockill, P and Brooks, A.~F. and Brown, D.~D. and Brunett, S and Buikema, A and Bulik, T and Bulten, H.~J. and Buonanno, A and Buskulic, D and Bustamante Rosell, M.~J. and Buy, C and Byer, R.~L. and Cabero, M and Cadonati, L and Cagnoli, G and Cahillane, C and Calder{\'{o}}n Bustillo, J and Callister, T.~A. and Calloni, E and Camp, J.~B. and Campbell, W.~A. and Canepa, M and Cannon, K.~C. and Cao, H and Cao, J and Capocasa, E and Carbognani, F and Caride, S and Carney, M.~F. and Carullo, G and Casanueva Diaz, J and Casentini, C and Caudill, S and Cavagli{\`{a}}, M and Cavalier, F and Cavalieri, R and Cella, G and Cerd{\'{a}}-Dur{\'{a}}n, P and Cerretani, G and Cesarini, E and Chaibi, O and Chakravarti, K and Chamberlin, S.~J. and Chan, M and Chao, S and Charlton, P and Chase, E.~A. and Chassande-Mottin, E and Chatterjee, D and Chaturvedi, M and Chatziioannou, K and Cheeseboro, B.~D. and Chen, H.~Y. and Chen, X and Chen, Y and Cheng, H.-P. and Cheong, C.~K. and Chia, H.~Y. and Chincarini, A and Chiummo, A and Cho, G and Cho, H.~S. and Cho, M and Christensen, N and Chu, Q and Chua, S and Chung, K.~W.},
    number = {3},
    month = {7},
    pages = {31040},
    volume = {9},
    doi = {10.1103/PhysRevX.9.031040},
    arxivId = {1811.12907}
}

@article{Aso2013InterferometerDetector,
    title = {{Interferometer design of the KAGRA gravitational wave detector}},
    year = {2013},
    journal = {Phys. Rev. D},
    author = {Aso, Yoichi and Michimura, Yuta and Somiya, Kentaro and Ando, Masaki and Miyakawa, Osamu and Sekiguchi, Takanori and Tatsumi, Daisuke and Yamamoto, Hiroaki},
    number = {4},
    month = {8},
    pages = {043007},
    volume = {88},
    publisher = {American Physical Society},
    url = {https://journals.aps.org/prd/abstract/10.1103/PhysRevD.88.043007},
    doi = {10.1103/PhysRevD.88.043007},
    issn = {15507998},
    arxivId = {1306.6747}
}

@article{Sedda2026IsolatedSources,
       author = {{Arca Sedda}, Manuel and {Paiella}, Lavinia and {Ugolini}, Cristiano and {Santoliquido}, Filippo and {Mestichelli}, Benedetta and {Usai}, Ilaria and {Simonato}, Filippo and {Branchesi}, Marica},
        title = "{Isolated or Dynamical? Tracing Black Hole Binary Formation through the Population of Gravitational-Wave Sources}",
      journal = {arXiv e-prints},
         year = 2026,
        month = mar,
          eid = {arXiv:2603.20430},
        pages = {arXiv:2603.20430},
          doi = {10.48550/arXiv.2603.20430},
archivePrefix = {arXiv},
       eprint = {2603.20430},
 primaryClass = {astro-ph.GA},
       adsurl = {https://ui.adsabs.harvard.edu/abs/2026arXiv260320430A}
}

@article{Uronen2026JointHoles,
       author = {{Uronen}, Laura and {Li}, Tian and {Janquart}, Justin and {Phurailatpam}, Hemanta and {Poon}, Jason and {Collett}, Thomas and {Koopmans}, Leon and {Hannuksela}, Otto},
        title = "{Joint Bayesian Source and Lens Reconstruction for Multi-messenger Binary Black Holes}",
      journal = {arXiv e-prints},
         year = 2026,
        month = mar,
          eid = {arXiv:2603.09794},
        pages = {arXiv:2603.09794},
          doi = {10.48550/arXiv.2603.09794},
archivePrefix = {arXiv},
       eprint = {2603.09794},
 primaryClass = {astro-ph.HE},
       adsurl = {https://ui.adsabs.harvard.edu/abs/2026arXiv260309794U}
}

@article{Birrer2018Lenstronomy:Package,
    title = {{lenstronomy: multi-purpose gravitational lens modelling software package}},
    year = {2018},
    author = {Birrer, Simon and Amara, Adam},
    url = {https://lenstronomy.readthedocs.io.},
    arxivId = {1803.09746v2}
}

@article{Phurailatpam2024LerSimulator,
    title = {{ler : LVK (LIGO-Virgo-KAGRA collaboration) event (compact-binary mergers) rate calculator and simulator}},
    year = {2024},
    journal = {arXiv e-prints},
    author = {Phurailatpam, Hemantakumar and More, Anupreeta and Narola, Harsh and Chung Yin, Ng and Janquart, Justin and Van Den Broeck, Chris and Akseli Hannuksela, Otto and Singh, Neha and Keitel, David},
    month = {7},
    pages = {arXiv:2407.07526},
    doi = {10.48550/arXiv.2407.07526},
    arxivId = {2407.07526}
}

@article{Ubach2025Self-lensingDegeneracy,
    title = {{Self-lensing of moving gravitational-wave sources can break the microlensing crossing timescale degeneracy}},
    year = {2025},
    author = {Ubach, Helena},
    month = {12},
    url = {https://arxiv.org/pdf/2512.08898},
    arxivId = {2512.08898}
}

\end{document}